\documentclass{phc-proc4-auth}
\usepackage{graphicx}
\usepackage{amsmath}
\usepackage{amssymb}

\begin{document}

\begin{frontmatter}

\title{Anisotropic properties
of MgB$_2$ by torque magnetometry}

\author[UNI,ETH]{M. Angst},
\ead{angst@physik.unizh.ch}
\author[UNI]{D. Di Castro},
\author[IFPAN]{R. Puzniak},
\author[IFPAN]{A. Wisniewski},
\author[ETH]{J. Jun},
\author[ETH]{S. M. Kazakov},
\author[ETH]{J. Karpinski},
\author[UNI]{S. Kohout},
\author[UNI]{H. Keller}

\address[UNI]{Physik-Institut, Universit\"at Z\"urich, 8057
Z\"urich, Switzerland}
\address[ETH]{Solid State Physics Laboratory ETH, 8093-Z\"urich, Switzerland}
\address[IFPAN]{Institute of Physics, Polish Academy of Sciences,
Aleja Lotnikow 32/46, 02-668 Warsaw, Poland}

\begin{abstract}
Anisotropic properties of superconducting MgB$_2$ obtained by
torque magnetometry are compared to theoretical predictions,
concentrating on two issues. Firstly, the angular dependence of
$H_{c2}$ is shown to deviate close to $T_c$ from the dependence
assumed by anisotropic Ginzburg-Landau theory. Secondly, from the
evaluation of torque vs angle curves it is concluded that the
anisotropy of the penetration depth $\gamma_ \lambda$ has to be
substantially higher at low temperature than theoretical
estimates, at least in fields higher than $0.2\,{\mathrm{T}}$.
\end{abstract}

\end{frontmatter}

Superconductivity in two bands of different dimensionality leads
to a temperature dependent anisotropy of the upper critical field
$\gamma _H = H_{c2}^{\|ab} / H_{c2}^{\|c}$ \cite{note1} in
MgB$_2$, observed, e.g., by torque magnetometry
\cite{Angst02PRL03PhyC}. Both torque results
\cite{Angst02PRL03PhyC} and calculations \cite{Miranovic03} of
$H_{c2}(\theta)$ indicate systematic deviations from the angular
dependence expected within anisotropic Ginzburg-Landau theory
(AGLT). However, the deviations found experimentally are most
pronounced near $T_c$, while the calculations \cite{Miranovic03}
predict pronounced deviations at low temperature $T$ only.
Recently, new calculations of $H_{c2}(\theta)$ were carried out
for the (intra-band) dirty limit \cite{Gurevich03,Golubov03}. We
will show that there is good agreement in the form of the
deviations of $H_{c2}(\theta)$ from AGLT between our torque
results and the calculations of Ref.\ \cite{Golubov03}.
Calculations \cite{Kogan02rapid} also predicted an anisotropy of
the penetration depth $\gamma _\lambda \ll \gamma _H$ at low $T$.
A field $H$ dependence of an effective anisotropy
\cite{Angst02PRL03PhyC} may be taken as an indication of such a
difference between $\gamma _\lambda$ and $\gamma _H$
\cite{Karpinski03SST}. A recent calculation of torque $\tau
(\theta)$ dependences in the London regime for the case of
different $\gamma _\lambda$ and $\gamma _H$ led to the prediction
of a sign reversal of the torque at low $T$ in MgB$_2$
\cite{Kogan02PRL}. From the comparison of the $\tau(\theta)$
dependence with the predictions of Ref.\ \cite{Kogan02PRL}, we
find a lower limit for $\gamma _\lambda$ at low temperatures,
considerably higher than theoretical estimates
\cite{Kogan02rapid}.

\begin{figure}[!tb]
\centering
\includegraphics[width=0.98\linewidth]{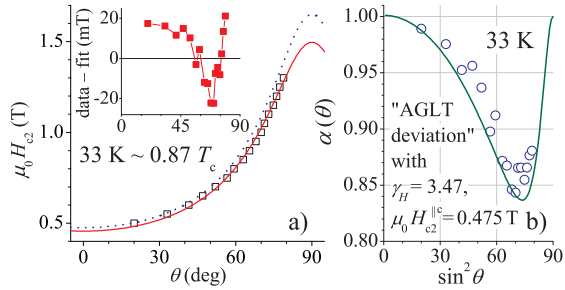}
\caption{ a) Upper critical field $H_{c2}$ vs angle $\theta$ of a
MgB$_2$ single crystal, at $0.87\,T_c$ (symbols). Free fit of AGLT
angular dependence (full line; $\mu _\circ
H_{c2}^{\|c}=0.456\,{\mathrm{T}}$, $\gamma _H = 3.28$) shows clear
systematic deviations, highlighted in the inset. AGLT dependence
with the same parameters as used in panel b) is also shown (dotted
line, $\mu _\circ H_{c2}^{\|c}=0.475\,{\mathrm{T}}$, $\gamma _H =
3.47$). b) "AGLT deviation" $\alpha(\theta)$ (see text) of the
data of panel a) (symbols) and a recent calculation
\cite{Golubov03} (full line).} \label{AGLTdeviations}
\end{figure}

For details concerning measurement apparatus and procedure,
samples, and the determination of $H_{c2}$ see Refs.\
\cite{Angst02PRL03PhyC,Karpinski03SST}. $H_{c2}(\theta)$,
determined from $\tau(\theta)$ curves measured in various fields
at $33\,{\mathrm{K}} \simeq 0.87\, T_c$, is shown in Fig.\
\ref{AGLTdeviations}a). By definition, $\tau$ is 0 for $H\|c$ or
$\|ab$, and small for field directions close. This is why there
are no data close to $0^{\circ}$ and $90^{\circ}$. In AGLT,
$H_{c2}(\theta)$ is described by
\begin{equation}
H_{\mathrm{c2}}^{\mathrm{AGL}}(\theta)=H_{\mathrm{c2}}^{\|c}  (
\cos^2 \theta + \sin^2 \theta /\gamma_H^2  ) ^{-1/2}.
\label{Hc2_theta}
\end{equation}
The best fit of Eq.\ (\ref{Hc2_theta}) to the data is indicated by
the full line. Small, but systematic deviations can be seen,
especially when plotting the difference between experimental data
and best fit vs $\theta$ (inset): at $0.87\, T_c$,
$H_{c2}(\theta)$ is {\em not} (accurately) described by Eq.\
(\ref{Hc2_theta}). Deviations from Eq.\ (\ref{Hc2_theta}) were not
observed at lower $T$ (cf.\ Fig.\ 2 of Ref.\
\cite{Angst02PRL03PhyC}). Deviations most pronounced in the region
of $0.9$-$0.95\, T_c$ were also found in a recent calculation
\cite{Golubov03} assuming high intraband scattering ({\em dirty
limit}). In order to compare experimentally observed deviations to
the predictions of Ref.\ \cite{Golubov03}, we calculated ``AGLT
deviations'' $\alpha (\theta) \equiv H_{c2} (\theta) /
H_{c2}^{\mathrm{AGL}} (\theta)$. For $\mu _\circ H_{c2}^{\|c}=
0.475\,{\mathrm{T}}$ and $\gamma = 3.47$, $\alpha (\theta)$ has
form and magnitude [Fig.\ \ref{AGLTdeviations}b)] very similar to
deviation functions for {\em calculated} \cite{Golubov03}
$H_{c2}(\theta)$ (full line) at the same temperature \cite{note2}.
Although the theoretically predicted \cite{Golubov03} $\gamma_H
\simeq 4.86$ is higher than our data indicate, the similarity of
the AGLT deviation suggests that (intraband) scattering cannot be
neglected in theoretical descriptions of $H_{c2}$.

\begin{figure}[!tb]
\centering
\includegraphics[width=0.95\linewidth]{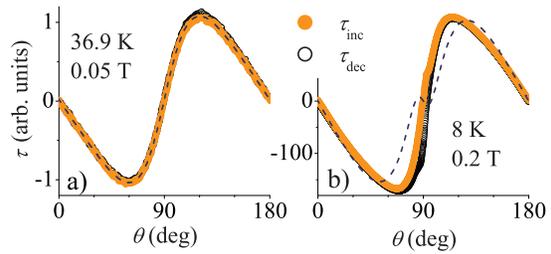}
\caption{ a) Torque $\tau$ vs angle $\theta$ measured close to
$T_c$. Dashed line: $\tau(\theta)$ calculated \cite{Kogan02PRL}
with $\gamma _\lambda = \gamma _H = 2$. b) $\tau$ vs $\theta$ at
low $T$. Dashed line: $\tau(\theta)$ calculated \cite{Kogan02PRL}
with $\gamma _\lambda = 2$, $\gamma _H = 6$.}
\label{tauvsangleTcomp}
\end{figure}

Figure \ref{tauvsangleTcomp}a) shows a $\tau(\theta)$ curve
measured (on a different crystal) in the mixed state close to $T_c
\simeq 38.5\,{\mathrm{K}}$. Near $T_c$, the difference between
$\gamma _\lambda$ and $\gamma _H$ is small, in agreement with
theoretical predictions \cite{Miranovic03,Kogan02PRL}. The
$\tau(\theta)$ curve measured at low $T$ [Fig.\
\ref{tauvsangleTcomp}b)] has the same sign as the one measured
close to $T_c$, i.e., there is no sign change as expected
\cite{Kogan02PRL} for $\gamma _\lambda \ll \gamma _H$. For $\gamma
_\lambda$ moderately lower than $\gamma _H$, Ref.\
\cite{Kogan02PRL} predicts a sign change only in an angular region
close to $90^{\circ}$, illustrated with a dashed line in Fig.\
\ref{tauvsangleTcomp}b). Such a partial sign change is also not
observed, the maximum angular region where it could occur given by
the irreversible region (the slight asymmetry in the
irreversibility is due to thickness variations of the crystal).
Comparing the data with curves calculated according to Ref.\
\cite{Kogan02PRL}, with $\mu _\circ H_{c2}^{\|c}=3\,{\mathrm{T}}$,
$\gamma_H=6$ \cite{Angst02PRL03PhyC} and various $\gamma
_\lambda$, we conclude that $\gamma _\lambda$ has to be at least
$2.6$, considerably higher than currently available theoretical
estimates \cite{Kogan02rapid}. Alternatively, if $\gamma _H$ in
$0.2\, {\mathrm{T}}$ is much smaller than in $H \approx H_{c2}$
\cite{note1}, the absence of a sign reversal is compatible with
smaller $\gamma _\lambda$. However, we should mention that the
best description of the data is given by $\gamma _\lambda \approx
\gamma _H \approx 3.3$.

The discrepancy may be explained by the influence of the magnetic
field, depressing superconductivity in the more isotropic $\pi$
bands. This should lead to anisotropies ($\gamma _\lambda$ and/or
$\gamma _H$) increasing with increasing field
\cite{Angst02PRL03PhyC}. An anisotropy increasing with $H$ has
also been postulated based on specific heat measurements (mostly
sensitive to the coherence length, i.e., $\gamma _H$ \cite{note1})
\cite{Bouquet02PRL}. Furthermore, recent neutron scattering
results indicate an increasing $\gamma _\lambda (H)$
\cite{EskildsenPC}. The calculations of Ref.\ \cite{Kogan02rapid}
are valid for the low field limit, which is difficult to
accurately probe due to irreversibility (no sign reversal was
found in $\tau(\theta)$ measured at low $T$ for
$0.03\,{\mathrm{T}} \! \leq \! \mu _\circ H \! \leq \! 1.5\,
{\mathrm{T}}$, but for the curves measured in the lowest fields,
it may be hidden by irreversibility).

In conclusion, deviations of the $H_{c2}(\theta)$ dependence from
the AGLT dependence near $T_c$ are well approximated by recent
calculations \cite{Golubov03}. While there is good agreement
between experiment and theory on $H_{c2}$, the penetration depth
anisotropy is still associated with open questions. Especially the
issue of the field influence on $\gamma _\lambda$ and $\gamma _H$
\cite{note1} deserves theoretical attention.

We thank A.~A. Golubov for providing calculated $H_{c2}(\theta)$
data for $0.87\, T_c$. This work was supported by the Swiss
National Science Foundation, the European Community
(ICA1-CT-2000-70018) and the Polish State Committee for Scientific
Research (5 P03B 12421).


\begin{thebibliography}{00}
\bibitem{note1}
For $H < H_{c2}$, $\gamma _H$ is generalized to mean the
anisotropy of the coherence lengths $\xi$.
\bibitem{note2} We note that the parameter values for which best agreement with
Ref.\ \cite{Golubov03} is seen are $4$-$5\%$ higher than those of
the AGLT ``best fit''.
\end{thebibliography}
\end{document}